\documentclass{PoS}

\title{Supernova remnants as cosmic ray factories}

\ShortTitle{Supernova remnants as cosmic ray factories}

\author{\speaker{Damiano Caprioli}%
        \\
       INAF-Osservatorio di Arcetri\\
       E-mail: \email{caprioli@arcetri.astro.it}}

\abstract{In this work we investigate particle acceleration in supernova remnant shocks within a semi-analytical formalism which self-consistently accounts for particle acceleration, amplification of the magnetic field via streaming instability and back-reaction of both accelerated particles and magnetic turbulence on the shock dynamics.
In particular, we study the interplay between particle injection and magnetic field amplification, showing how a phenomenological but reasonable saturation of the standard streaming instability may lead to quite steep spectra for the accelerated particles.
We comment on the implications that such a scenario may have on the comprehension of the diffuse spectrum of Galactic cosmic rays and of gamma-ray observations of single SNRs.}

\FullConference{25th Texas Symposium on Relativistic Astrophysics \\
		 December 6-10, 2010\\
		 Heidelberg, Germany}

\begin{document}

\section{The standard scenario}
The association between supernova remnants (SNRs) and Galactic cosmic rays (CRs) has been very popular since 1934, when Baade and Zwicky argued that this class of astrophysical objects can account for the required CR energetics \cite{baade-zwicky34}.
Along with this argument, the so-called SNR paradigm for the origin of CRs has been supported also by E. Fermi's proposal of a very general mechanism for particle acceleration, which turned out to be particularly efficient for particles diffusing around SNR forward shocks \cite{bell78a,blandford-ostriker78}.
This diffusive shock acceleration (DSA) mechanism is also very appealing because it naturally returns power-law spectra $\phi(E)\propto E^{-\Gamma}$ for the energy distribution of accelerated particles, with $\Gamma=2$ for large sonic Mach numbers.
Moreover, such a spectral index does not depend on the details of the scattering processes but only on the compression ratio felt by diffusing particles.

\subsection{The diffuse spectrum of Galactic CRs}
By measuring the ratios between secondary and primary CRs, it is possible to infer that CRs diffuse during their propagation in the Milky Way and that the Galactic residence time scales as $E^{-\delta}$, with $\delta\simeq 0.3-0.6$.
Since the CR spectrum measured at Earth is $\propto E^{-2.7}$ from about a GeV up to the \emph{knee} (around $3\times 10^{6}$GeV), in this wide range of energies we have a strong observational constraint, namely $\Gamma+\delta\simeq 2.7$ and, therefore, $\Gamma=2.1-2.4$.
CR spectra at the sources may thus be steeper than the ones predicted by standard DSA at SNR forward shocks \cite{hillas05}. 

This point is made even more problematic by the fact that, in order to account for correct energetics, SNR shocks have to channel large fractions (from 10\% to 50\%, depending on the Galactic propagation details) of their kinetic energy into accelerated particles. 
Therefore, the SNR paradigm for the origin of CRs naturally suggests SNR shocks to be modified by the dynamical back-reaction of particles they accelerate: any reliable DSA theory has to be non-linear (NLDSA), i.e.\ has to include the feedback of CR pressure and energy in the shock hydrodynamics.

The most important output of the kinetic NLDSA theory, independently of the approach adopted for describing the coupling between CR transport and SNR hydrodynamics \cite{comparison}, is that the spectrum of accelerated particles has to be concave, i.e. steeper (harder) than $E^{-2}$ ---the standard prediction of DSA--- at low (high) energies, as a consequence of the shock modification induced by an efficient particle acceleration \cite{drury83,jones-ellison91,malkov-drury01}.  
This prediction leads to some very natural questions: why such a concavity is not observed in the diffuse spectrum of Galactic CRs? And also, how can NLDSA be consistent with source spectra steeper than $E^{-2}$ up to the knee?

Before trying to answer these questions it is necessary to stress that, in principle, there is no one-to-one correlation between the \emph{instantaneous} spectrum of particles being accelerated at a shock, whose slope is predicted by Fermi's mechanism, and the spectrum of particles released from a SNR during its lifetime.
The latter is in fact the convolution over different evolutionary stages of (concave?) spectra, with different normalizations and cut-offs. 
In particular, during the Sedov-Taylor stage the instantaneous maximum energy decreases with time as a consequence of the decrease of both shock velocity and magnetization level: at any given time, the highest-energy particles cannot but escape the system from upstream \cite{escape}.  
A very simple argument \cite{crspectrum} demonstrates how the self-similar adiabatic evolution of a SNR during its Sedov-Taylor stage may also lead to a $\sim E^{-2}$ spectrum for these escaping particles.
The actual details of how and when accelerated particles are released into the Galaxy and become CRs are still poorly understood.
Different possibilities for particle escape are investigated in Ref.~\cite{crspectrum}, where it is also shown that, most likely, escape from upstream is necessary to account for the highest-energy part of the broadband spectrum, while at lower energies, say below 1 TeV, the source spectrum is almost invariably dominated by the accelerated particles which have been advected downstream, suffered adiabatic losses, and finally released at the ``death'' of the remnant itself.   

In all of the NLDSA calculations of the most abundant CR species \cite{bv07,za10,nuclei} the spectral slope at the source has never been found to be larger than $\Gamma\simeq 2.1-2.15$, in turn implying $\delta\geq 0.55$. 
Nevertheless, as discussed e.g. in \S 7 of Ref.~\cite{hillas05}, a strong energy dependence in the residence time is at odds with the low level of anisotropy measured in the CR arrival direction, which seems to point towards the lowest limit consistent with the measured B/C ratio, namely $\delta\simeq 0.3$.

\subsection{Hadronic emission from SNRs}
There is another independent piece of information coming from the present generation of $\gamma$-ray telescopes, which are providing us with an unprecedented wealth of data from sources coincident with SNRs.
Very briefly, there are two main scenarios in principle able to account for the high-energy photons detected in the GeV band by Fermi and AGILE and in the TeV band by MAGIC, VERITAS, HESS, CANGAROO, MILAGRO, etc.: the leptonic scenario, in which $\gamma$-rays are due to inverse Compton scattering or bremsstrahlung by relativistic electrons, or the hadronic scenario, in which they are due to the decay of neutral pions produced in nuclear collisions between the relativistic nuclei and the background plasma \cite{dav94}.
Even if there are no clear-cut evidences of the predominance of a scenario over the other yet, we cannot but notice that basically all of the SNRs detected in $\gamma$-rays show a photon spectrum softer than $E^{-2}$ \cite{gamma}. 

This fact is particularly intriguing because inverse Compton scattering returns rather hard photon spectra, while both relativistic bremsstrahlung and pion decay provide a one-to-one map of the parent electron and proton spectra in terms of spectral slope: it is very likely that by observing $\gamma$-ray bright SNRs we can probe the instantaneous spectrum of accelerated particles and, in turn, have an insight into how NLDSA effectively works.

In this respect, it is interesting to notice that most of the detected SNRs show spectra in the range $E^{-2.2}-E^{-2.4}$, in nice agreement with the adoption of a mildly energy-dependent Galactic residence time and with the measured anisotropy. 
Putting all these evidences together, one may conclude that the most natural acceleration mechanism at work in SNRs has to produce spectra $\phi(E)\propto E^{-2.3}$, hence significantly steeper than what predicted by standard NLDSA calculations. 

\section{CRs and magnetic field amplification}
We want to exploit the semi-analytical formalism put forward in Refs.~\cite{boundary, crspectrum} to investigate the possibility to obtain such a steep source spectrum in the context of a NLDSA theory by including also the effects induced by a very efficient magnetic field amplification (MFA).
The strong non-thermal X-ray luminosity of young SNRs has been in fact explained as due to synchrotron emission by relativistic electrons radiating in magnetic fields as high as a few hundred $\mu$G, i.e.\ orders of magnitudes larger than the typical interstellar ones (e.g.\ \cite{jumpl} and references therein).

These strong magnetic fields are very likely produced by the same accelerated particles via plasma instabilities which, in the quasi-linear regime, saturate when the ratio between the pressure in accelerated particles and in magnetic turbulence is of order $P_{cr}/P_{B}\simeq 2 M_{A}$, where $M_{A}=V_{sh}/V_{A}$ is the Alfv\'enic Mach number of the shock \cite{bell78a}. 
We extrapolate this result to the non-linear regime $\delta B\gg B_{0}$ by assuming that the relevant Alfv\'en velocity has to be calculated in the amplified magnetic field $\delta B$, and not in the background one as prescribed by the quasi-linear theory.
Additional comments and \emph{caveats} related to this choice can be found e.g.\ in Refs.~\cite{gamma,jumpkin}.

Very generally, there are several effects induced by a very efficient MFA: 
1) the pressure in magnetic turbulence may easily become dominant with respect to the thermal plasma pressure, governing the hydrodynamics and preventing an excessive shock modification \cite{jumpkin,veb07};
2) the velocity of the waves particles scatter against, which is of order $\sim V_{A}$, may become a non-negligible fraction of the fluid velocity, altering in a significant way the compression ratios actually felt by CRs and eventually making their spectrum steeper than $E^{-2}$ \cite{bell78a,jumpkin,gamma};
3) since $M_{A}(\delta B)\ll M_{A}(B_{0})$, for a given gradient in the CR pressure the instability saturation might be achieved for larger $\delta B$.

\begin{figure} 
\includegraphics[width=.5\textwidth, height=5cm]{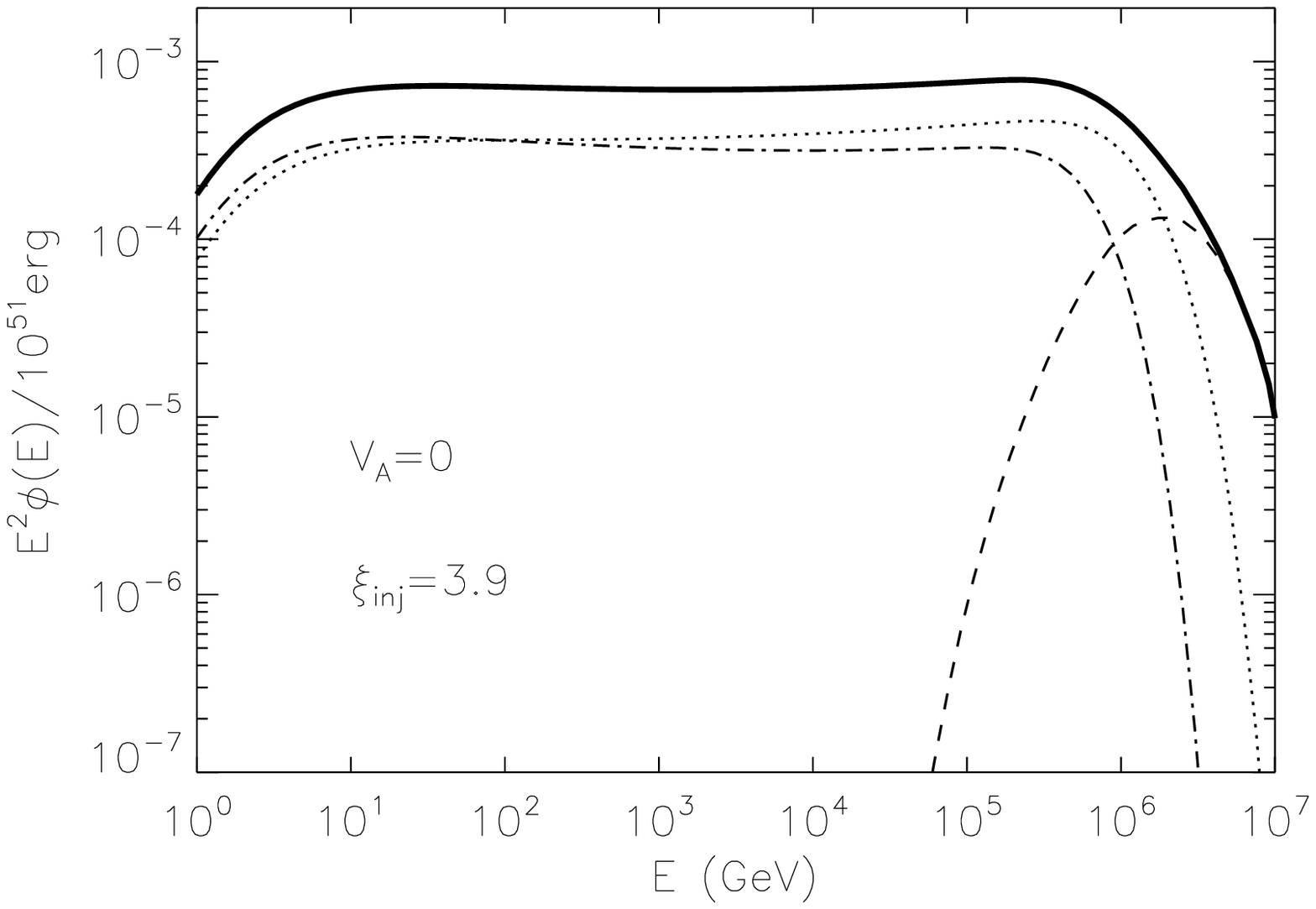}
\includegraphics[width=.5\textwidth, height=5cm]{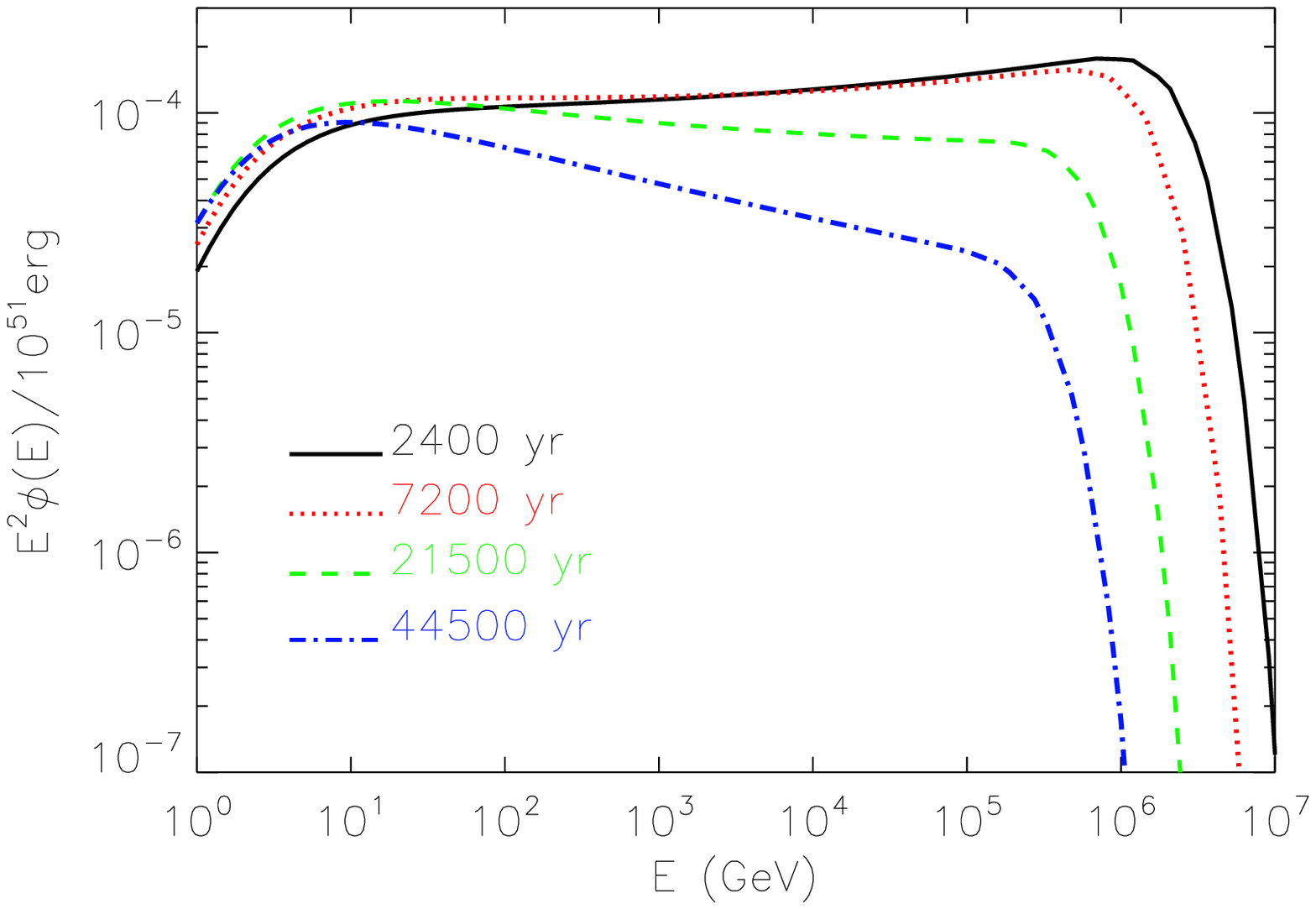} 
\includegraphics[width=.5\textwidth, height=5cm]{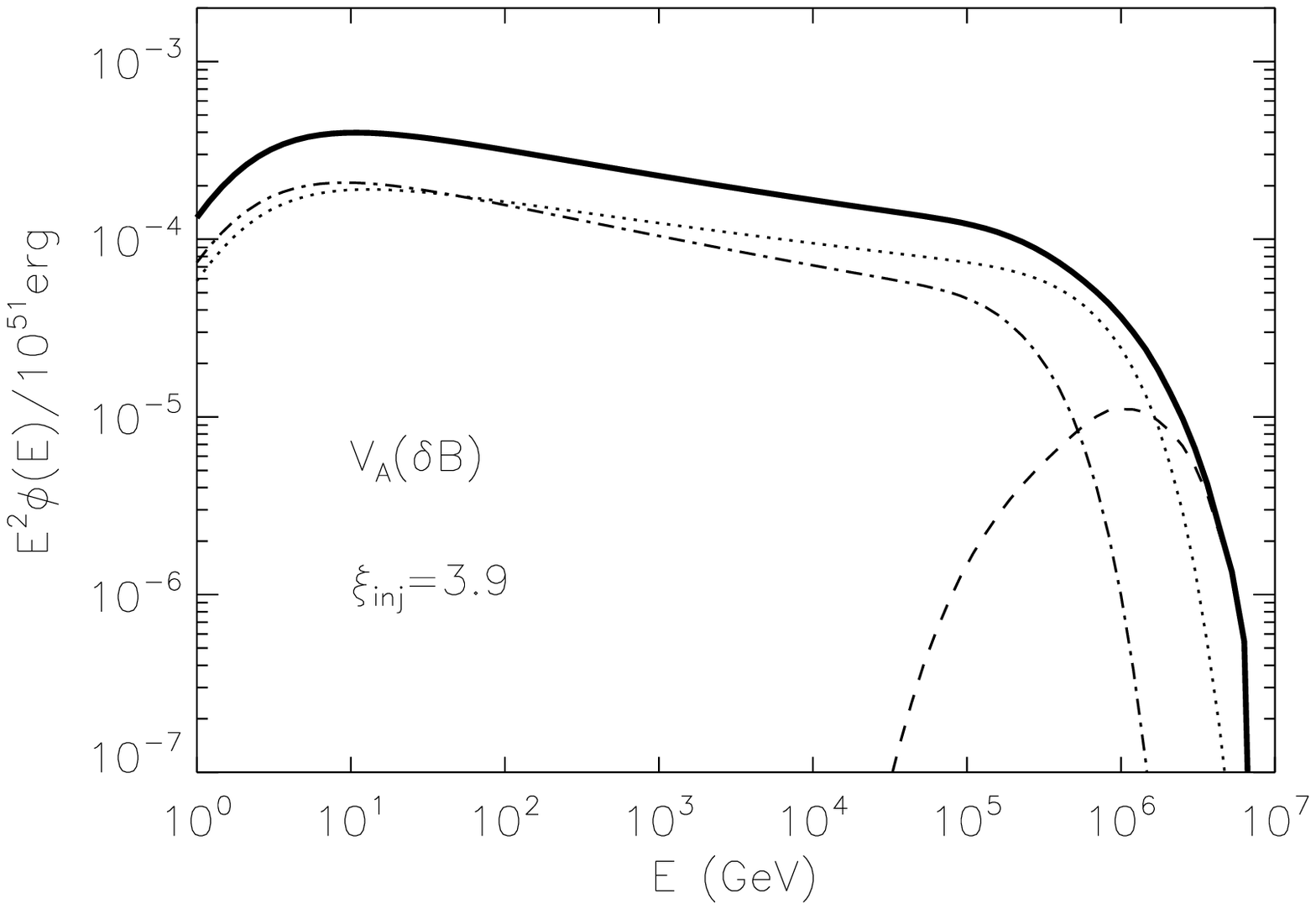} 
\includegraphics[width=.5\textwidth, height=5cm]{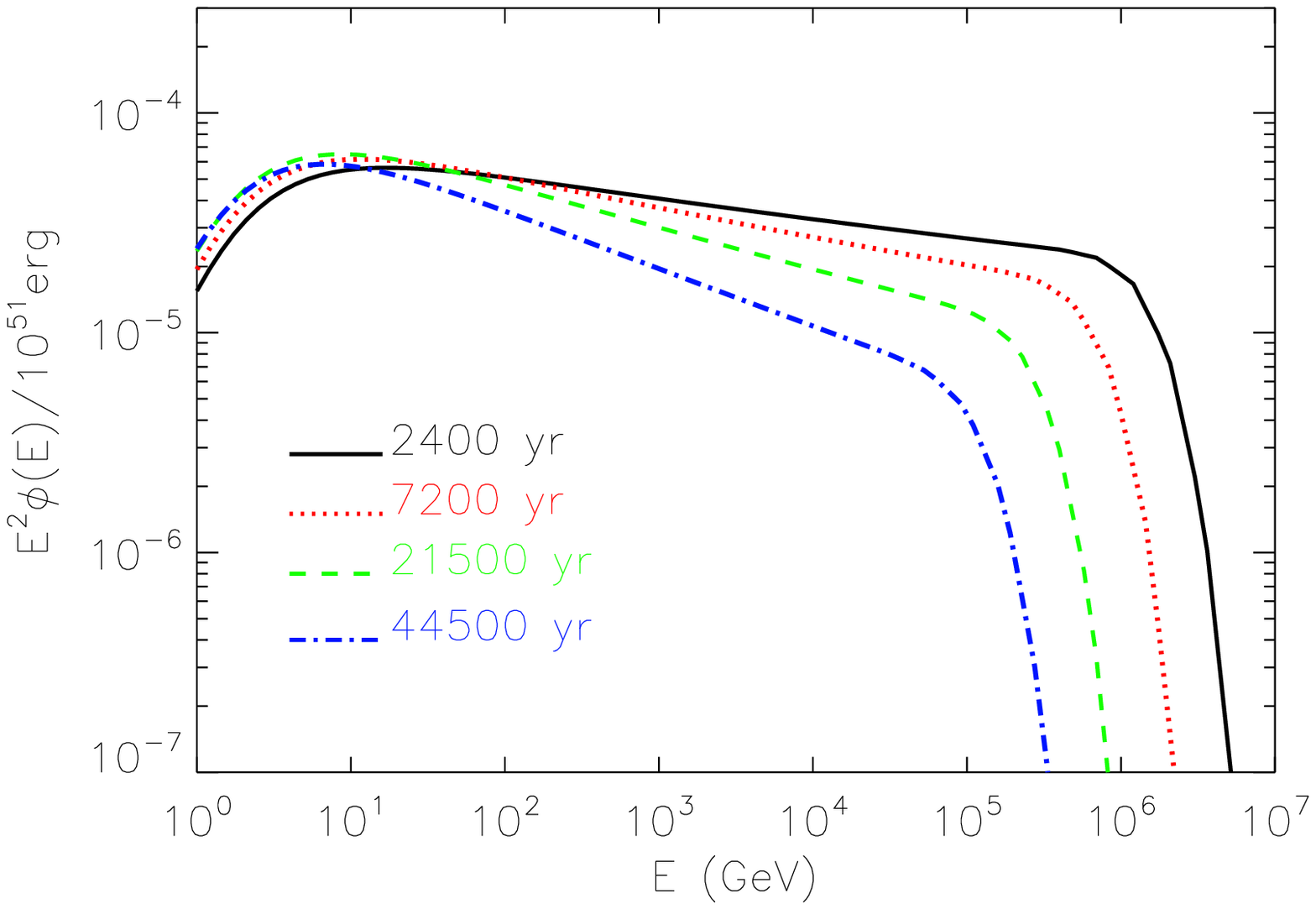}
\includegraphics[width=.5\textwidth, height=5cm]{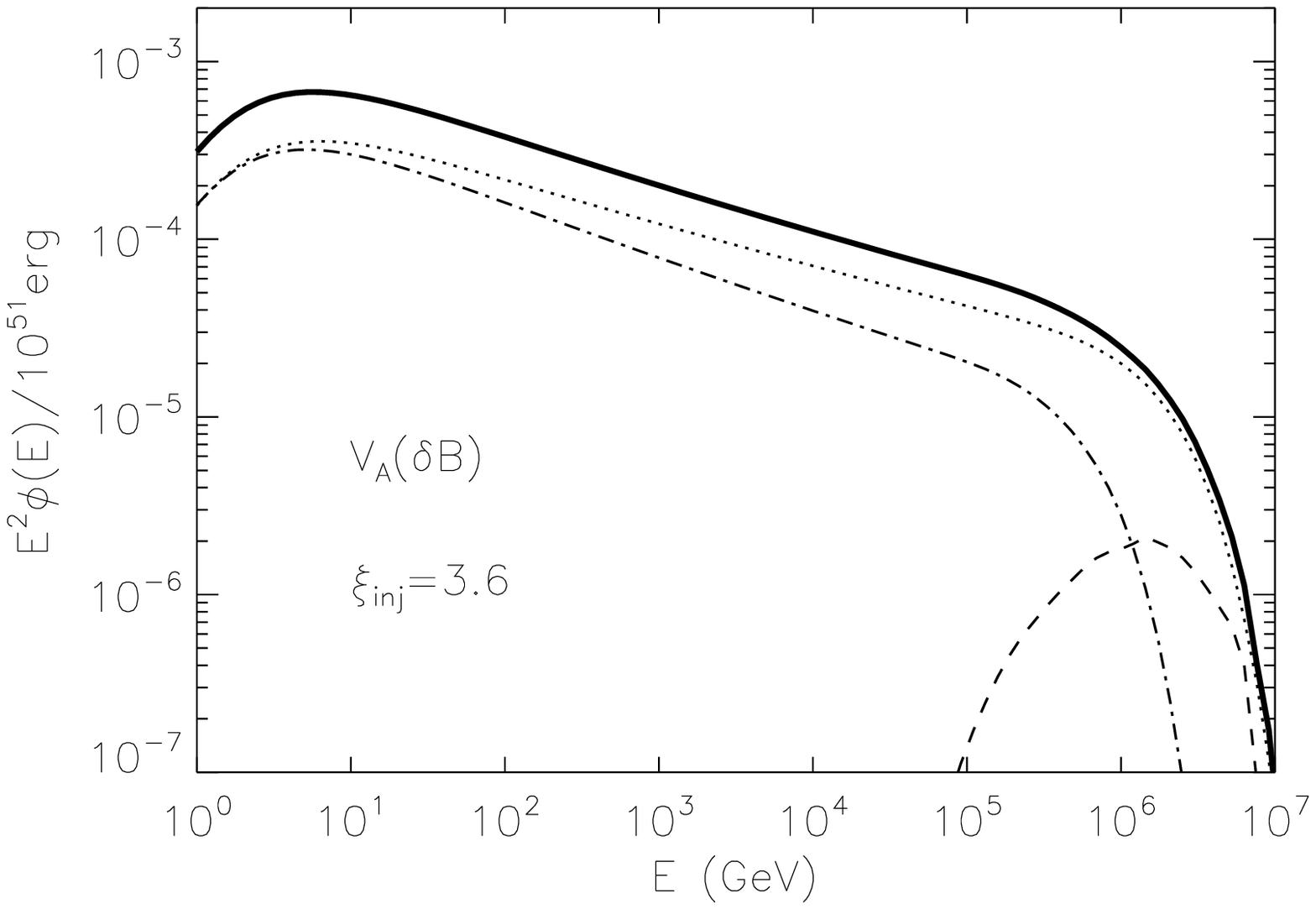}
\includegraphics[width=.5\textwidth, height=5cm]{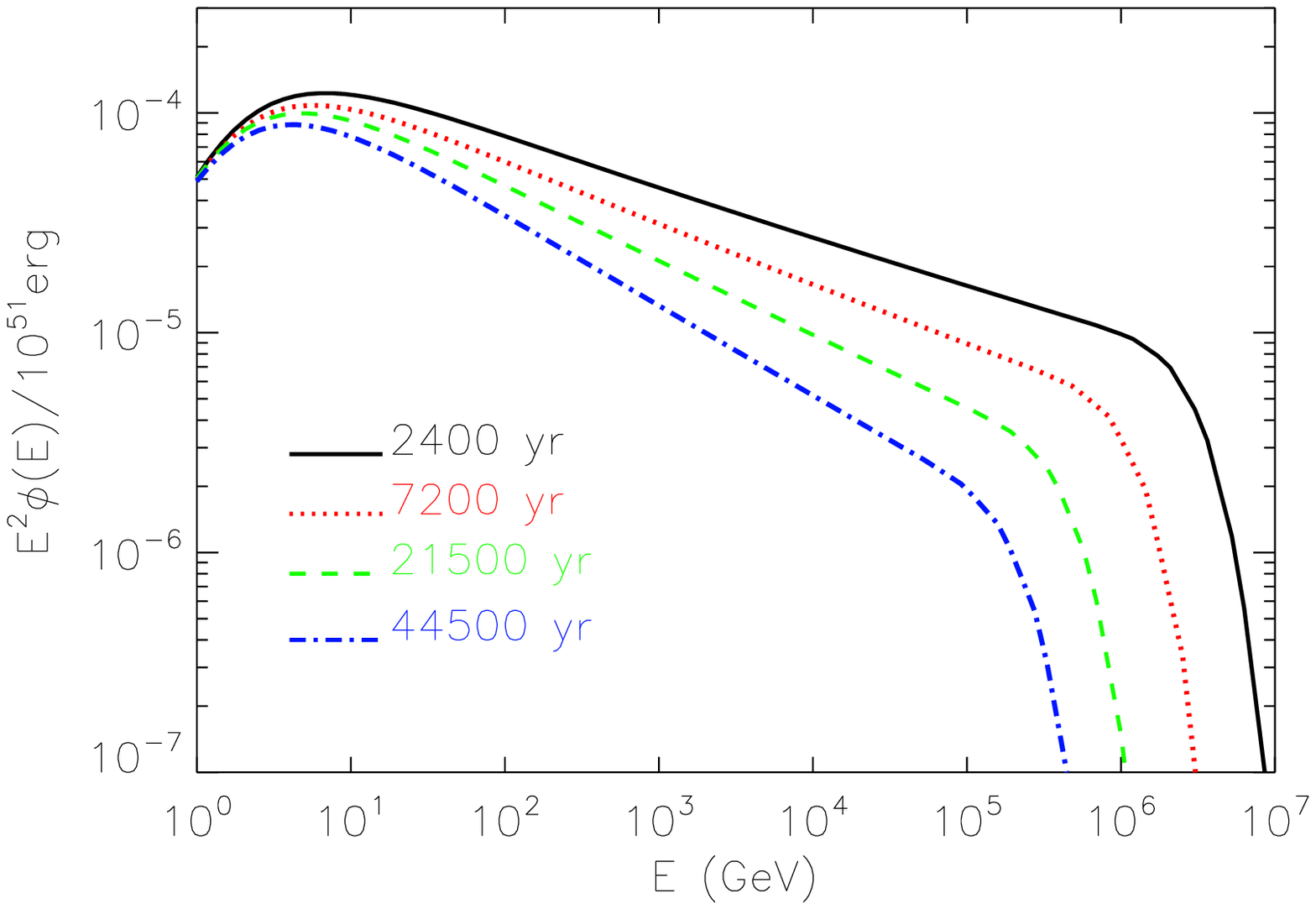} 
\includegraphics[width=.5\textwidth, height=5cm]{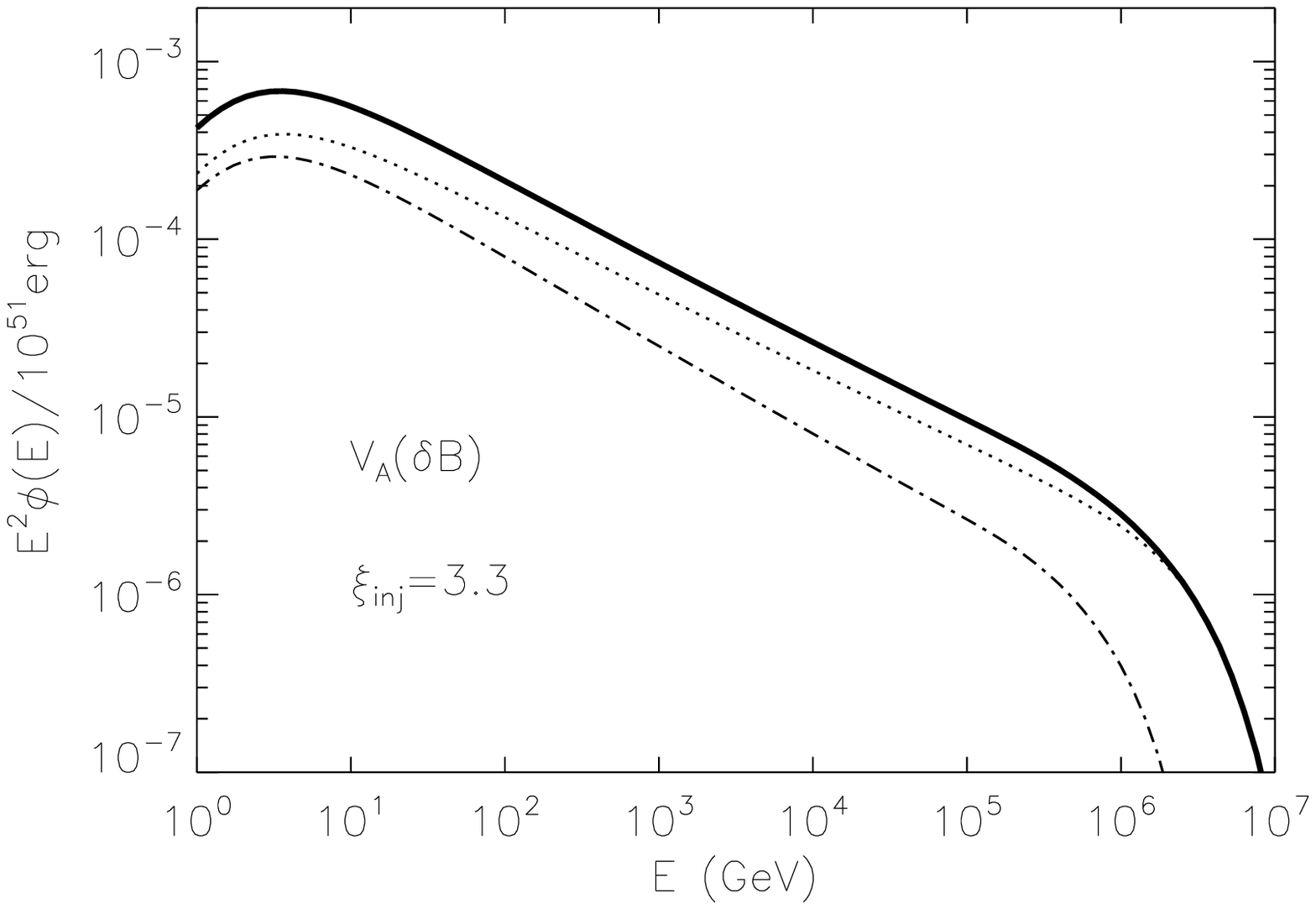}
\includegraphics[width=.5\textwidth, height=5cm]{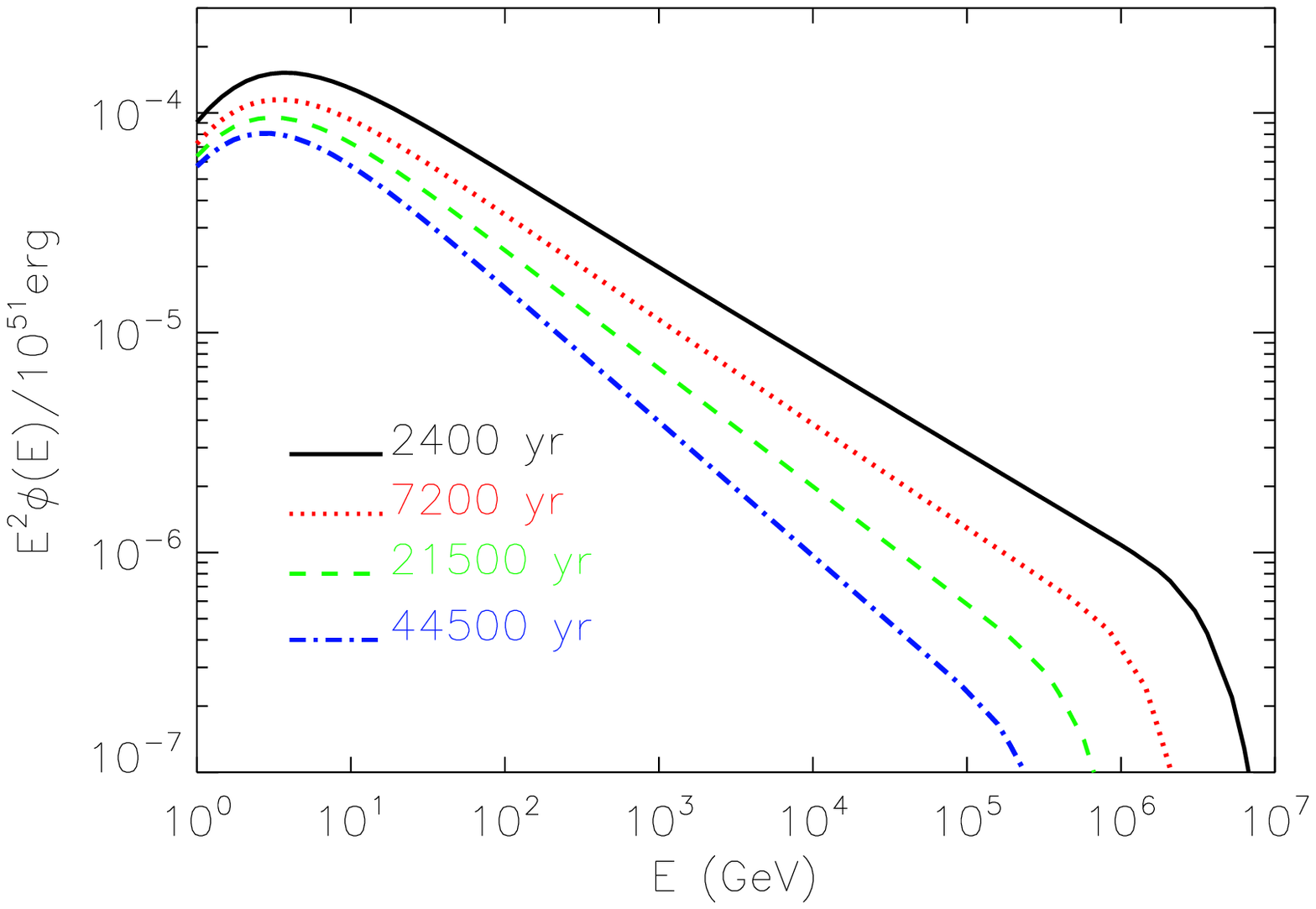} 
\caption{Source spectra without ($V_{A}=0$, top panels) and with the Alfv\'enic drift included ($V_{A}(\delta B)$, other panels), for different injection efficiencies $\xi_{\rm inj}=3.9,3.6,3.3$, as in the legend.
The solid lines in the left panels represent the total CR spectrum, distinguished into contributions due to escape from upstream (dashed), leakage from downstream (dotted) and release at the SNR death (dot-dashed lines). In the right panels instantaneous spectra of accelerated particles are instead shown as a function of the SNR age.
 } 
\label{fig1} 
\end{figure} 

We consider a SN explosion releasing $1.4$ solar masses and $10^{51}$erg in the interstellar medium with density, temperature and magnetic field equal to 0.01 protons cm$^{-3}$, $10^{6}${\rm K} and 5$\mu$G, as in Ref.~\cite{crspectrum}.
The self-similar shock evolution is coupled with a non-linear treatment of particle acceleration, shock dynamics and MFA, including particle injection from the thermal bath \cite{BGV05}.
In particular, all the protons with momentum $\xi_{\rm inj}$ times larger than the thermal momentum downstream are assumed to be injected into the acceleration process (thermal leakage): the smaller $\xi_{\rm inj}$, the larger the fraction of injected particles.

MFA amplification upstream is taken as due to resonant streaming instability, producing a magnetic field given by
\begin{equation}\label{satur}
\frac{\delta B^{2}}{8\pi}=P_{B}\simeq\frac{P_{cr}}{2M_{A}}.
\end{equation}
The dynamical back-reaction of such an amplified magnetic field on the shock dynamics is also included \cite{jumpl,jumpkin}. 
Particle escape due to the decrease of the SNR confining power during the Sedov-Taylor stage is accounted for by imposing the CR distribution function to vanish beyond a free-escape boundary placed at a distance 0.5$R_{sh}$ upstream of the shock \cite{escape,boundary}, while escape from downstream is treated as in Ref.~\cite{crspectrum} by posing $\beta=0.1$.

In the left panels of Fig.~\ref{fig1} the total CR spectra produced by our benchmark remnant are shown (thick solid lines) as the sum of three different contributions: escape from upstream (dashed lines), leakage from downstream (dotted) and release at the end of the Sedov-Taylor stage, when the SNR non-thermal activity is expected to fade (dot-dashed).
In the right panels, instead, some snapshots of instantaneous proton spectra are shown; the first age corresponds to the transition from the ejecta-dominated to the adiabatic stage, when energies as high as the knee are achieved \cite{BAC07}.

We consider here two different configurations for the velocity of the scattering centers, in order to illustrate how crucial this ingredient may be. 
The top panels of Fig.~\ref{fig1} show a case in which MFA is at work but scattering centers comove with the fluid, i.e.\ $V_{A}=0$. 
This case is compared with the panels below, where $V_{A}$ is included and calculated in the amplified magnetic field $\delta B$, instead. 

There are two very important points to be noticed.
First, considering an effective Alfv\'enic drift in the amplified magnetic field leads to spectra much steeper than in the case $V_{A}=0$, where the convolution over time of concave spectra leads to a power-law spectrum slightly flatter than $E^{-2}$, as already shown in Ref.~\cite{crspectrum}.

Second, when the Alfv\'en velocity (and hence the saturation of the resonant streaming instability) are taken in the amplified magnetic field, total spectra are straight power-laws whose spectral indexes correlate with the parameter which tunes the thermal leakage efficiency, $\xi_{\rm inj}$, here taken to vary between 3.3 and 3.9.
In particular, second, third and fourth rows in Fig.~\ref{fig1} have increasingly larger thermal leakages, corresponding to inject into the acceleration process fractions as large as $\sim 10^{-6}$ ($\xi_{\rm inj}=3.9$), $\sim 10^{-5}$ ($\xi_{\rm inj}=3.6$) and $\sim 10^{-4}$ ($\xi_{\rm inj}=3.3$) of the particles crossing the shock.
We can interpret this behavior as follows: the larger the number of particles injected, the larger the CR pressure and, as a consequence of Eq.~\ref{satur}, the larger the self-generated magnetic turbulence.
This last process, however, produces a crucial non-linear feedback which makes the particle spectrum steeper and eventually reduces the amount of energy channelled into CRs. 
The balance between these two opposite effects returns self-consistent solutions with both efficient particle acceleration (around 10\%) and, occasionly, very steep spectra, like the case $\xi_{\rm inj}=3.3$ in Fig.~\ref{fig1}.
The spectral slopes of the total SNR spectra are in fact $\Gamma\simeq 2.15, 2.28$ and 2.45 for the cases with $V_{A}(\delta B)$.   
The fine structure of such a non-linear interplay will be widely discussed in a forthcoming paper.

\section{Conclusions}
The details of how MFA occurs in SNR shocks may radically change our previous comprehension of efficient  DSA theories, leading to a revised interpretation of many observational facts.
In order to illustrate this point, we calculated instantaneous and total CR spectra produced at a typical SNR forward shock by adopting a rigorous semi-analytical kinetic approach to NLDSA, on top of which we added a phenomenological description of very efficient Alfv\'enic turbulence generation due to resonant streaming instability. 
 
We showed that it is possible to account both for quite large acceleration efficiencies (around 10\%) and, at the same time, for CR spectra significantly steeper than what predicted by previous NLDSA calculations.
More precisely, when the Alfv\'en velocity, $V_{A}$, is calculated in the amplified magnetic field, the intrinsic non-linear interplay between accelerated particles, magnetic field amplification and shock dynamics provides steeper and steeper spectra for larger and larger number of injected particles, actually reversing the trend one would obtain by taking $V_{A}$ in the background magnetic field, as prescribed by quasi-linear theory.

In terms of slope, normalization and cut-off the spectra we obtained are consitent with basically all the observational evidences inferred from the diffuse spectrum of Galactic CRs and from $\gamma$-ray-brigth SNRs \cite{gamma}.

\bibliographystyle{JHEP}
\bibliography{bibtexas}

\end{document}